\documentclass[
    ,final            
  ]
  {aipproc}

\usepackage{amsmath} 
\usepackage{amssymb}
\usepackage{psfrag}

\layoutstyle{6x9}

\begin{document}

\newcommand{\zbar}{\bar{z}}
\newcommand{\odd}{\mathbb{O}}
\newcommand{\pom}{\mathbb{P}}
\newcommand{\tmin}{t_{\rm min}}
\newcommand{\smin}{s_{\rm min}}
\newcommand{\gev}{{\rm GeV}}

\hspace{10.5cm}\begin{minipage}[r]{4cm}
LPT-ORSAY 08-86\\
CPHT-PC 082.1008
\end{minipage}

\title{Pomeron-Odderon interference in production of $\pi^+\pi^-$ pairs in 
ultraperipheral collisions}


\author{B. Pire}{
  address={CPhT, \'Ecole Polytechnique, CNRS, 91128 Palaiseau, France}
}

\author{F. Schwennsen}{
  address={LPT, Universit\'e d'Orsay, CNRS, 91404 Orsay, France},
  altaddress={CPhT, \'Ecole Polytechnique, CNRS, 91128 Palaiseau, France}
}
\author{L. Szymanowski}{
  address={Soltan Institute for Nuclear Studies, Warsaw, Poland}
}

\author{S. Wallon}{
  address={LPT, Universit\'e d'Orsay, CNRS, 91404 Orsay, France}
}

\begin{abstract}
In this contribution we discuss the production of two pion pairs in high energy photon collisions as they can be produced in ultraperipheral collisions at hadron colliders such as the Tevatron, RHIC or LHC. 
We find that charge asymmetries may reveal the existence of the perturbative Odderon. 
\end{abstract}

\maketitle


\section{Introduction}

At high energies amplitudes of hadronic reactions with rapidity gaps are dominated by the exchange of a color singlet, $C$-even state  -- called the Pomeron. In the language of perturbative QCD 
 the Pomeron can be described at lowest order as the exchange of two gluons in the color singlet state. In contrast to the very well settled notion of the Pomeron, the status of its $C$-odd partner -- the Odderon -- is less safe. Although it is needed {\it e.g.} to describe properly  the different behaviors of $pp$ and $\bar p p$ elastic cross sections \cite{LN}, it still evades confirmation in the perturbative regime, where, again at lowest order, it can be described by the exchange of three gluons in the color singlet state.

The difficulty is rooted in the smaller amplitude for Odderon exchange in comparison to the Pomeron exchange. Hence,  in cross sections after squaring the amplitude, the Odderon contribution is always covered by the Pomeron. In this contribution we study charge asymmetries in the  production of two pion pairs in photon-photon collisions
\begin{equation}  
\gamma (q)\;\; \gamma (q') \to \pi^+(p_+)\;\; \pi^-(p_-)\;\;
\pi^+(p'_+)\;\; \pi^-(p'_-)\;, \label{gg}
\end{equation}
In such asymmetries, due to interference effects, the Odderon amplitude enters linearly and not quadratically the observable. This approach has been initiated in Ref.~\cite{Brodsky:1999mz}. In our specific case we consider the momentum transfer $t = (q - p_+ -p_-)^2$ to provide a hard scale of a few $\gev^2$ justifying a perturbative calculation within $k_T$-factorization since at the same time we impose $s\gg |t|$.

\section{Kinematics, Amplitudes and GDAs}

\begin{figure}[t]
\centerline{\includegraphics[height=5.5cm]{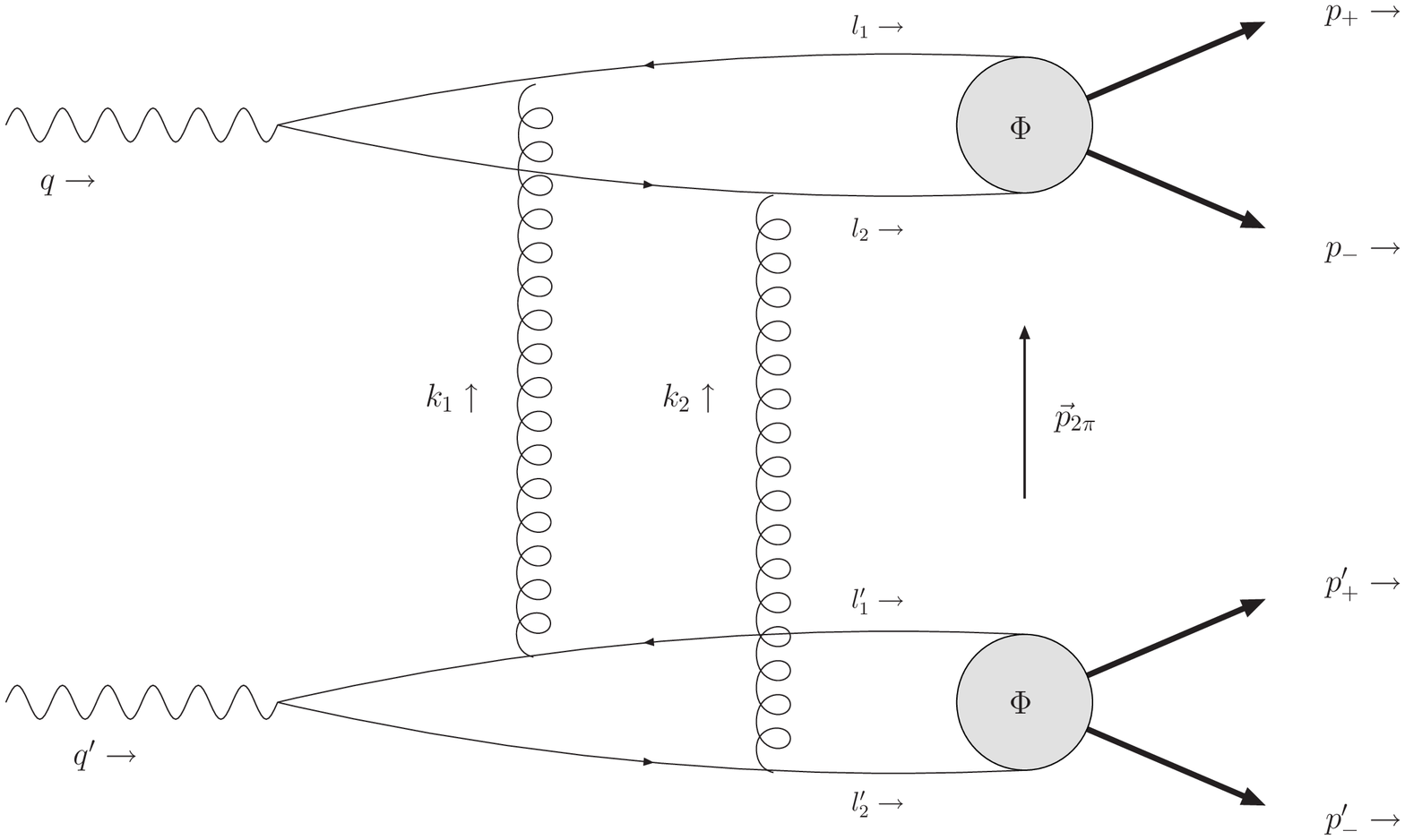}}
\caption{{\protect\small Kinematics of the reaction $\gamma \gamma \to \pi^+ \pi^-\;\; \pi^+ \pi^-$ in a sample Feynman diagram of the two gluon exchange process.}}
\label{fig:1}
\end{figure}

A sample diagram of the two gluon exchange is given in  Fig.~\ref{fig:1}.
Due to high energy factorization, the amplitudes can be expressed as convolutions of two impact factors over the transverse momenta of the exchanged gluons. The impact factors themselves consist of a perturbatively calculable part -- describing the transition of a photon into a quark-antiquark pair -- and a non-perturbative
part, the two pion generalized distribution amplitude (GDA) which parametrize the quark-antiquark to hadron transition. 

One key point to our final predictions is the choice of the phenomenological input: the GDA \cite{Hagler:2002nh,Hagler:2002sg,Warkentin:2007su} which are functions of the longitudinal momentum fraction $z$ of the quark,  of
the angle $\theta$ (in the same rest frame of the pion pair) and of the invariant mass $m_{2\pi}$ of the pion system. After an expansion in Gegenbauer polynomials $C_n^m(2z-1)$ and in Legendre polynomials $P_l(\beta\cos\theta)$ (where $\beta=\sqrt{1-4m_\pi^2/m_{2\pi}^2}$) \cite{Polyakov:1998ze}, it is believed that only the first terms give a significant contribution:
\begin{eqnarray}
  \Phi^{I=1} (z,\theta,m_{2\pi}) &=& 6z\zbar\beta f_1(m_{2\pi}) \cos\theta ,\\
  \Phi^{I=0} (z,\theta,m_{2\pi}) &=& 5z\zbar(z-\zbar)\left[-\frac{3-\beta^2}{2}f_0(m_{2\pi})+\beta^2f_2(m_{2\pi})P_2(\cos\theta)\right],
\end{eqnarray}
where $f_1(m_{2\pi})$ can be identified with the electromagnetic pion form factor $F_\pi(m_{2\pi})$. 
For the $I=0$ component we use different models. The first model follows Ref.~\cite{Hagler:2002nh} and expresses the functions $f_{0/2}$ in terms of the Breit-Wigner amplitudes of the according resonances.
A second model has been elaborated in Ref.~\cite{Warkentin:2007su} and interprets the functions $f_{0/2}$ as corresponding Omn\`es functions for $S-$ and $D-$waves constructed by dispersion relations from the phase shifts of the elastic pion scattering.

It has been argued \cite{Warkentin:2007su,Ananthanarayan:2004xy} that the actual phases of the GDA might be closer to the phases of the corresponding $T$ matrix elements $\frac{\eta_l e^{2i\delta_l}-1}{2i}$. The third model for the $I=0$ component of the GDA takes this into account by using the technique of model 2 with these phases $\delta_{T,l}$ of the $T$ matrix elements. Indeed, measurements at HERMES \cite{Airapetian:2004sy} do not observe a resonance effect at the $f_0$-mass, but concerning the $f_2$ both phases ($\delta_2$ and $\delta_{T,2}$) are compatible with data \cite{Warkentin:2007su}. Having this in mind, we consider also a fourth model -- a mixed description with the $f_0$ contribution from model 3 and the $f_2$ contribution from model 2. 

\section{Charge Asymmetries}

The GDAs for $C$-even pion pairs ($\Phi^{I=0}$) enter the Odderon exchange amplitude, while those for the $C$-odd pion pairs ($\Phi^{I=1}$) enter the Pomeron exchange. They are orthogonal to each other in  the space of Legendre polynomials in $\cos\theta$ such that only the interference term survives, when the amplitude squared is multiplied by $\cos\theta$ before the angular integration. Thereby we define a charge asymmetry in the following way:
\begin{eqnarray}
  &&A(t,m_{2\pi}^2,m_{2\pi}'^2) = \frac{\int\cos\theta\,\cos\theta'\,d\sigma(t,m_{2\pi}^2,m_{2\pi}'^2,\theta,\theta')}{\int\,d\sigma(t,m_{2\pi}^2,m_{2\pi}'^2,\theta,\theta')} \nonumber\\
 &=& \frac{\int_{-1}^1d\cos\theta\int_{-1}^1d\cos\theta'\;2\cos\theta\,\cos\theta'\,{\rm Re}\left[\mathcal{M}_\pom(\mathcal{M}_\odd+\mathcal{M}_{\gamma})^*\right]}{\int_{-1}^1d\cos\theta\int_{-1}^1d\cos\theta'\,\left[\left|\mathcal{M}_\pom\right|^2+\left|\mathcal{M}_\odd+\mathcal{M}_{\gamma}\right|^2\right]} ,
\end{eqnarray}
where also the $C$-odd photon exchange has been included. Since in the kinematic region of interest it is much smaller than the Odderon contribution, the asymmetry is driven by the Odderon/ Pomeron-interference.

The obtained landscape as a function of the two invariant masses would be difficult to measure.
To reduce the complexity, we integrate over the invariant mass of one of the two pion systems to obtain
\begin{eqnarray}
  \hat{A}(t,m_{2\pi}^2;m_{\rm min}^2,m_{\rm max}^2) &=& \frac{\int_{m_{\rm min}^2}^{m_{\rm max}^2} dm_{2\pi}'^2\int\cos\theta\,\cos\theta'\,d\sigma(t,m_{2\pi}^2,m_{2\pi}'^2,\theta,\theta')}{\int_{m_{\rm min}^2}^{m_{\rm max}^2} dm_{2\pi}'^2\int\,d\sigma(t,m_{2\pi}^2,m_{2\pi}'^2,\theta,\theta')}
. \label{eq:ahat}
\end{eqnarray}

An analytic calculation of the Odderon matrix element would demand the notion of analytic results for two-loop box diagrams, whose off-shellness for all external legs is different. With the techniques available on the market such a calculation is beyond the scope of this work. Instead we rely on a numerical evaluation by Monte Carlo methods. In particular we make use of a modified  version of {\sc Vegas} as it is provided by the {\sc Cuba} library \cite{Hahn:2004fe}.

\begin{figure}
  \centering
  \psfrag{A}{\footnotesize\!\!\!\!$A$}
  \psfrag{m2pi1}{\footnotesize\!\!\!\!\!\!\!\!\!\!\!\!\!\!\!\!\!$m_{2\pi}$ [GeV]}
  \psfrag{m2pi2}{\vspace{10mm}\footnotesize $m_{2\pi}'$ [GeV]}
  \includegraphics[width=5.2cm]{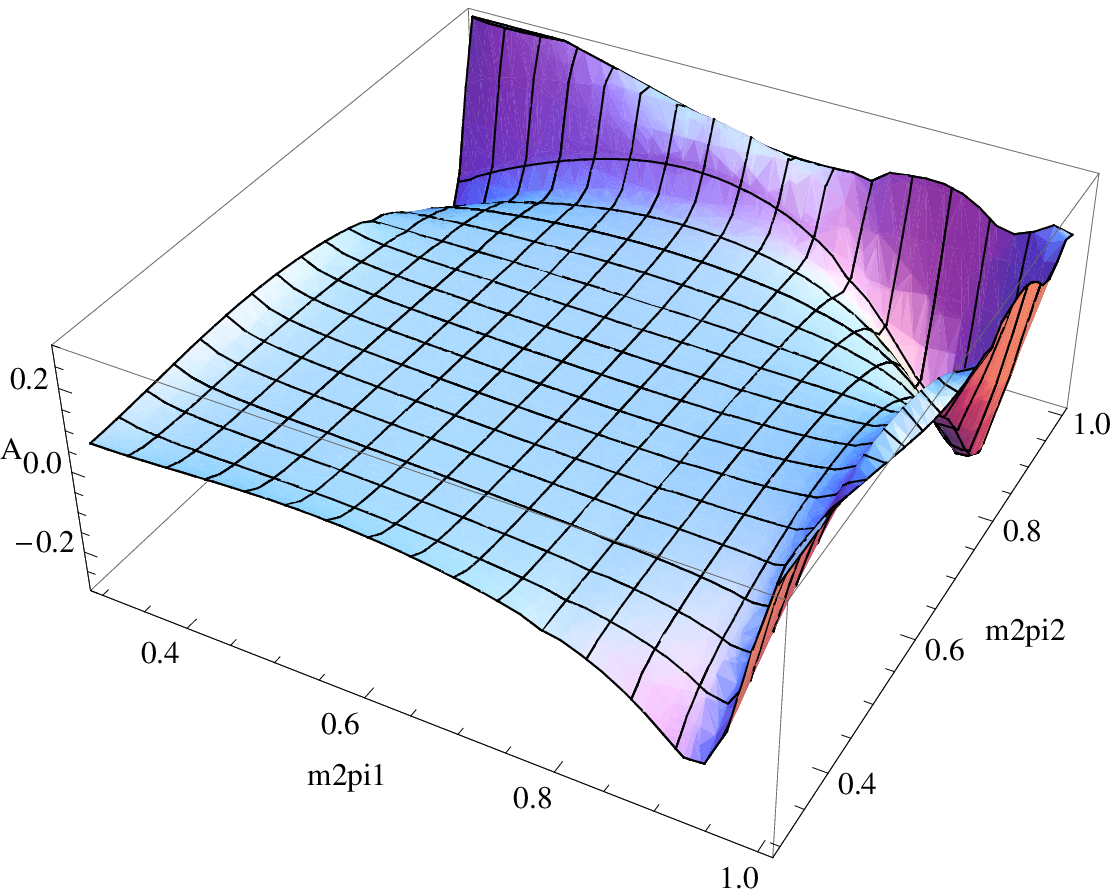}\hspace{15mm}
  \includegraphics[width=5.2cm]{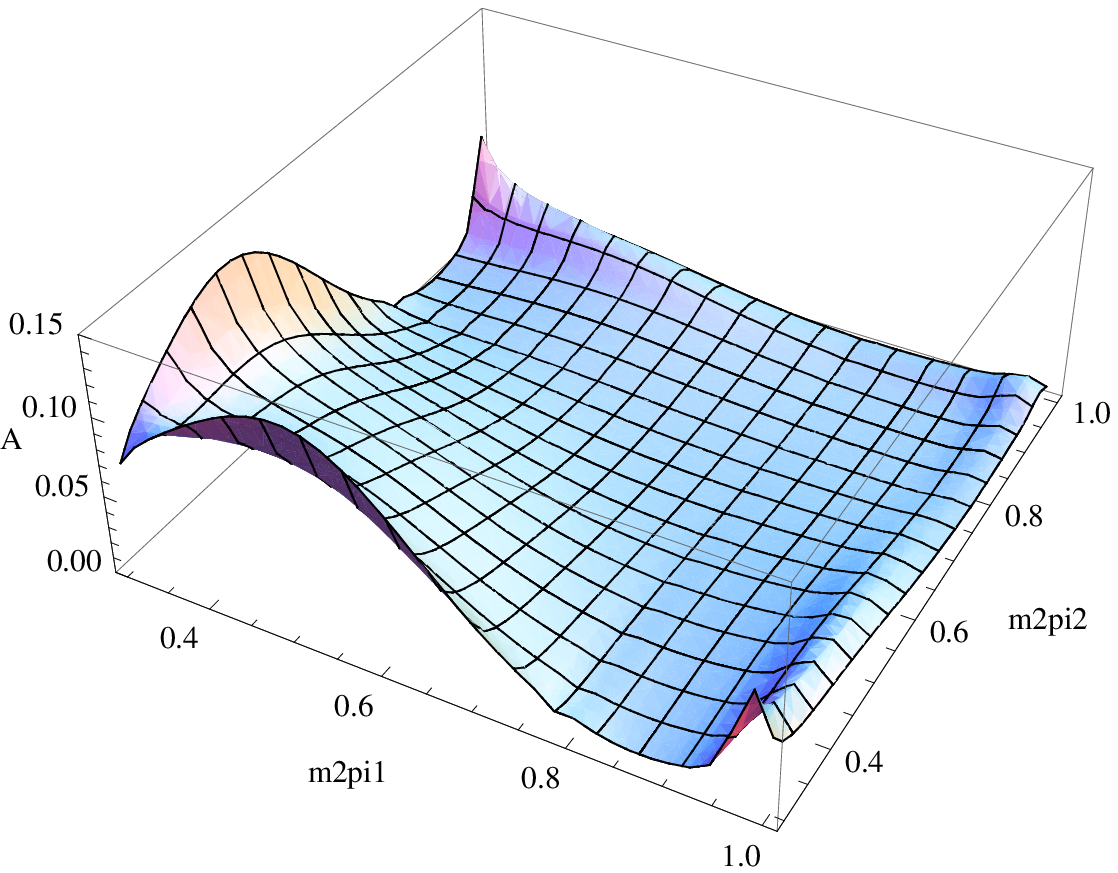}
  \label{fig:landscape}
  \caption{Asymmetry $A$ at $t=-1\,\gev^2$ for model 2 (left) and 3 (right). The shape of model 1 is very similar to model 2, and that of model 4 very similar to model 3.}
\end{figure}

\begin{figure}
  \centering
  \includegraphics[width=5.5cm]{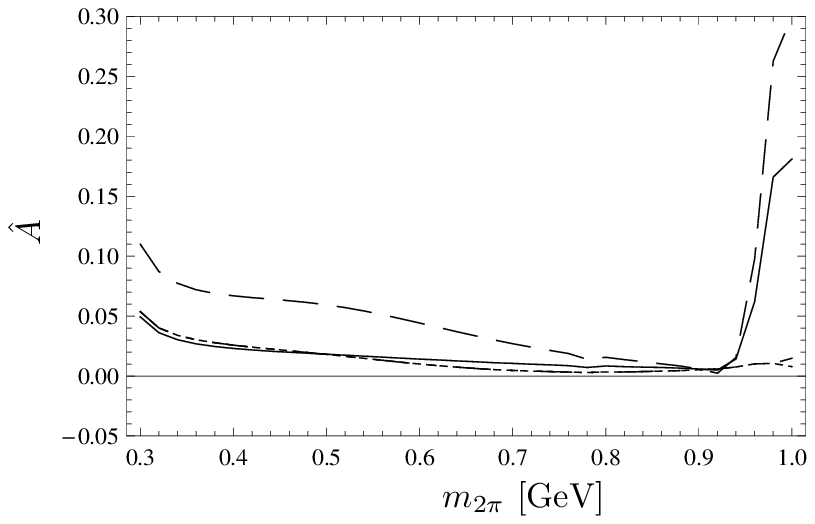}\hspace{10mm}
  \includegraphics[width=5.5cm]{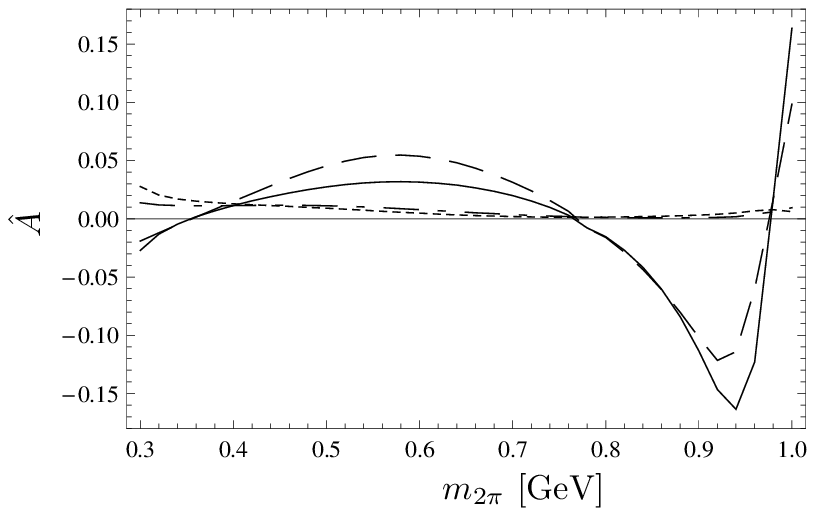}
  \caption{Asymmetry $\hat{A}$ at $t=-1\,\gev^2$ for model 1 (solid), 2 (dashed), 3 (dotted), and 4 (dash-dotted) -- model 3 and 4 are nearly on top of each other. Left column has $m_{\rm min}=.3\,\gev$ and $m_{\rm max}=m_\rho$, while right column has $m_{\rm min}=m_\rho$ and $m_{\rm max}=1\,\gev$. 
}
\label{fig:asymplot1}
\end{figure}

Although the asymmetry $A$ will problably not be measured, it is illustrative to display it for completeness in Fig.~\ref{fig:landscape}.
The result for the asymmetry $\hat{A}$ at $t=-1\,\gev^2$  is shown in Fig.~\ref{fig:asymplot1}. Since our framework is only justified for $m_{2\pi} ^2 < -t$, (in fact strictly speaking, one even needs $m_{2\pi}^2 \ll -t$ ), we keep $m_{2\pi}$ below 1\,GeV.

\section{Conclusion}

We have presented charge asymmetry estimates in production of pion pairs in $\gamma\gamma$ collisions. This asymmetry is linearly dependent on the Odderon amplitude and moreover is sizable but GDA-model dependent. HERMES measurements of two pion  electroproduction 
\cite{Airapetian:2004sy} disfavor models with a strong $f_0$ coupling to  
the $\pi^+ \pi^-$ state but to our minds higher statistics data, which  
may come from a JLab experiment at 6 or 12\,GeV, are needed  
before  a definite conclusion.
As we argue in Ref.~\cite{Pire:2008xe}, in $pp$ collisions at the LHC one can expect of the order of $10^3$ events per year. While the rates at RHIC would be far too low, at Tevatron a first search could be possible.

\begin{theacknowledgments}
We acknowledge discussions with Mike Albrow, Gerhard Baur, David d'Enterria, Bruno Espagnon, and Rainer Schicker.
This work is supported in part by the Polish Grant N N202 249235, the French-Polish scientific agreement Polonium, by the grant ANR-06-JCJC-0084 and by the ECO-NET program, contract 12584QK.
\end{theacknowledgments}

\bibliographystyle{aipproc}   


\end{document}